\documentclass[epj]{svjour}
\usepackage{graphicx}
\begin{document}
\title{Different amplitude and time distribution of the sound of light and classical music}
\author{P. Diodati\inst{1,2} \and S. Piazza\inst{1,2,}
\thanks{\email {stefano.piazza@pg.infn.it}}%
}                     
%
%
\institute{Dipartimento di Fisica dell'Universit\'{a} degli Studi di Perugia, Via Pascoli, 06123 Perugia, Italy \and Istituto Nazionale per la Fisica della Materia - Sezione di Perugia, Via Pascoli, 06123 Perugia, Italy}
\date{Received: date / Revised version: date}
%
\abstract{
Several pieces of different musical kinds were studied measuring $N(A)$, the output amplitude of a peak detector driven by the electric signal arriving to the loudspeaker. Fixed a suitable threshold $\bar{A}$, we considered $N(A)$, the number of times that $A(t)>\bar{A}$, each of them we named event and $N(t)$, the distribution of times $t$ between two consecutive events. Some $N(A)$ and $N(t)$ distributions are displayed in the reported logarithmic plots, showing that jazz, pop, rock and other popular rhythms have noise-distribution, while classical pieces of music are characterized by more complex statistics. We pointed out the extraordinary case of the aria ``\textit{La calunnia \`{e} un venticello}'', where the words describe an avalanche or seismic process, calumny, and the rossinian music shows $N(A)$ and $N(t)$ distribution typical of earthquakes.
} 
\maketitle
\section{Introduction}
\label{intro}
	It is well known that it is possible to establish a correspondence between sounds and numbers: given a composition we can digitize it no matter how complex it is. So we can characterize a piece of music completely by a set of numbers, and distinguish it from another in a mathematical way even if the two pieces appear identical even to the ears of good musicians \cite{pierce}. \par
This line of thinking runs contrary to the beliefs of the past and even to
day some experts are of the same opinion, believing that ``music is not as tangible or measurable as a liquid or a solid'' \cite{bent}. It is possible to characterize an elementary sound as well as a complex combination of sounds, even in presence of non linear effects. Eventually, what is not ``tangible or measurable'' are the listener's impressions caused by music. In principle we could measure the effect of music on the mind, but such measurement would be affected by culture, sensitiveness and emotivity of the listener, and therefore not universal. In this way we have a proof of the ``relativity'' of the judgment, influenced by a sort of musical ``common sense'' that takes shape during infancy and adolescence. So, in physics we have the common sense and in music we have the musical taste, formed by the music listened to from birth.\par
It is well known that the first attempts to find the mathematical hall-mark of some authors, with the aim to disclose the secret of the ``good music'', were made studying the distribution and repetition of notes. With the coming of computers, complex statistical studies can be made \cite{camill}, independent of the listener's judgment. Now it is possible to play electronic music while a program writes the score and shows it on the video. The succession of notes could be seen as a stochastic process. A Beethoven piece, for example, is able to be represented as a stochastic sequence (Markov's chain) where each event (each note) has an exactly-calculable recurrence probability\cite{bent}. With this kind of statistical study new descriptions and characterizations are possible, showing that universal features are not dependent on the listener's aesthetic judgment.

\section{Statistical analysis}
\label{sec:1}
Statistical study was performed considering $V^{2}(t)$, where $V(t)$ is the audio signal, as the function associated with music. Voss and Clarke \cite{voss1,voss2} found that the spectral density $S(f)$ of $V^{2}(t)$ exhibits a $1/f$ power spectra in the low frequency range $(<10 Hz)$ in many musical selections and in English speech. Recent measurements 
\cite{boon,nett} give 
$S(f) \simeq  f^{-\nu}$ in the low frequency range with $\nu \neq 1$, but no correlation between the analyzed music and the $\nu$ value seems to emerge.\par
\begin{figure}[!tbh]
\begin{center}
\includegraphics[trim=50 100 20 100,width=0.40\textwidth,keepaspectratio]{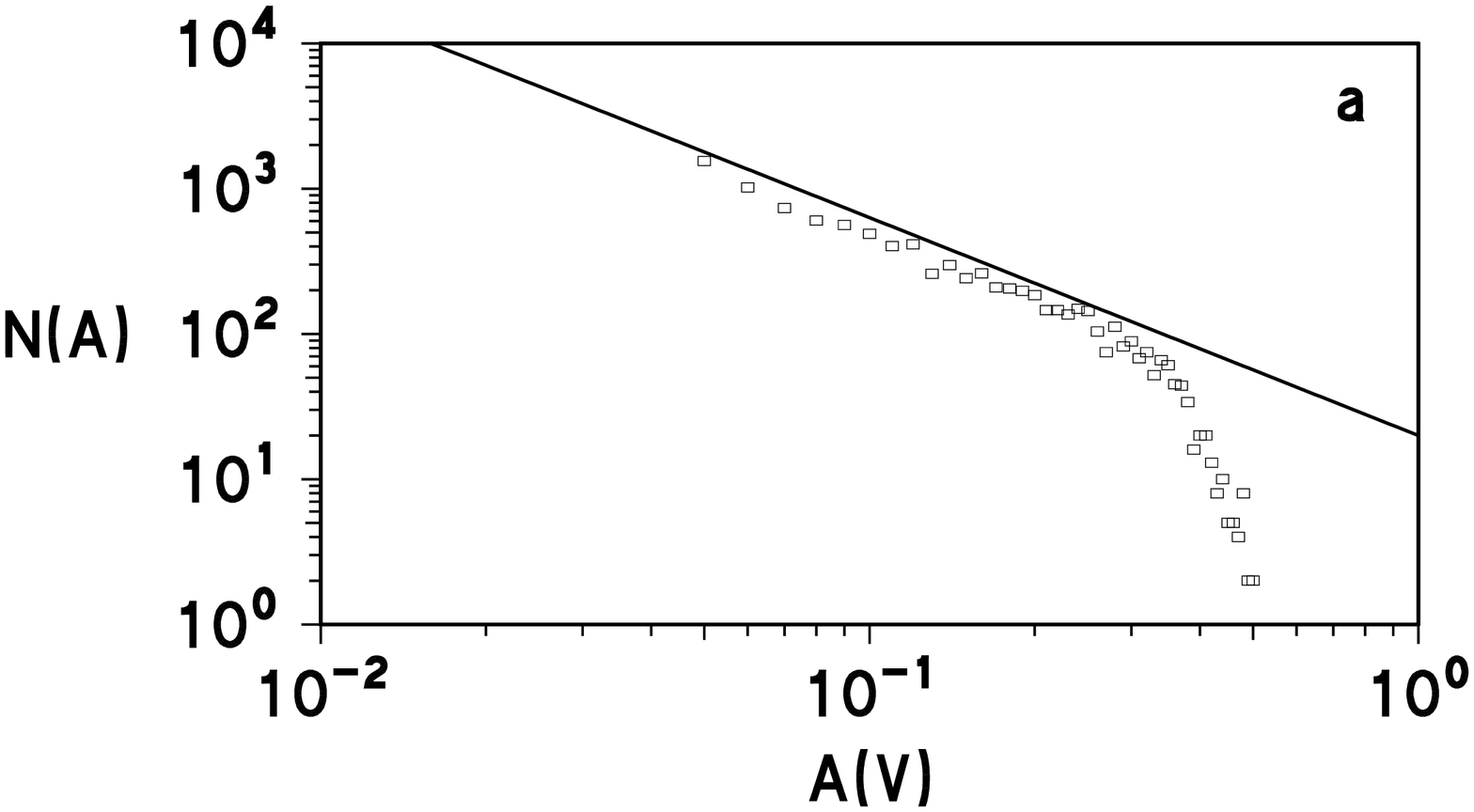}
\includegraphics[trim=50 100 20 100,width=0.40\textwidth,keepaspectratio]{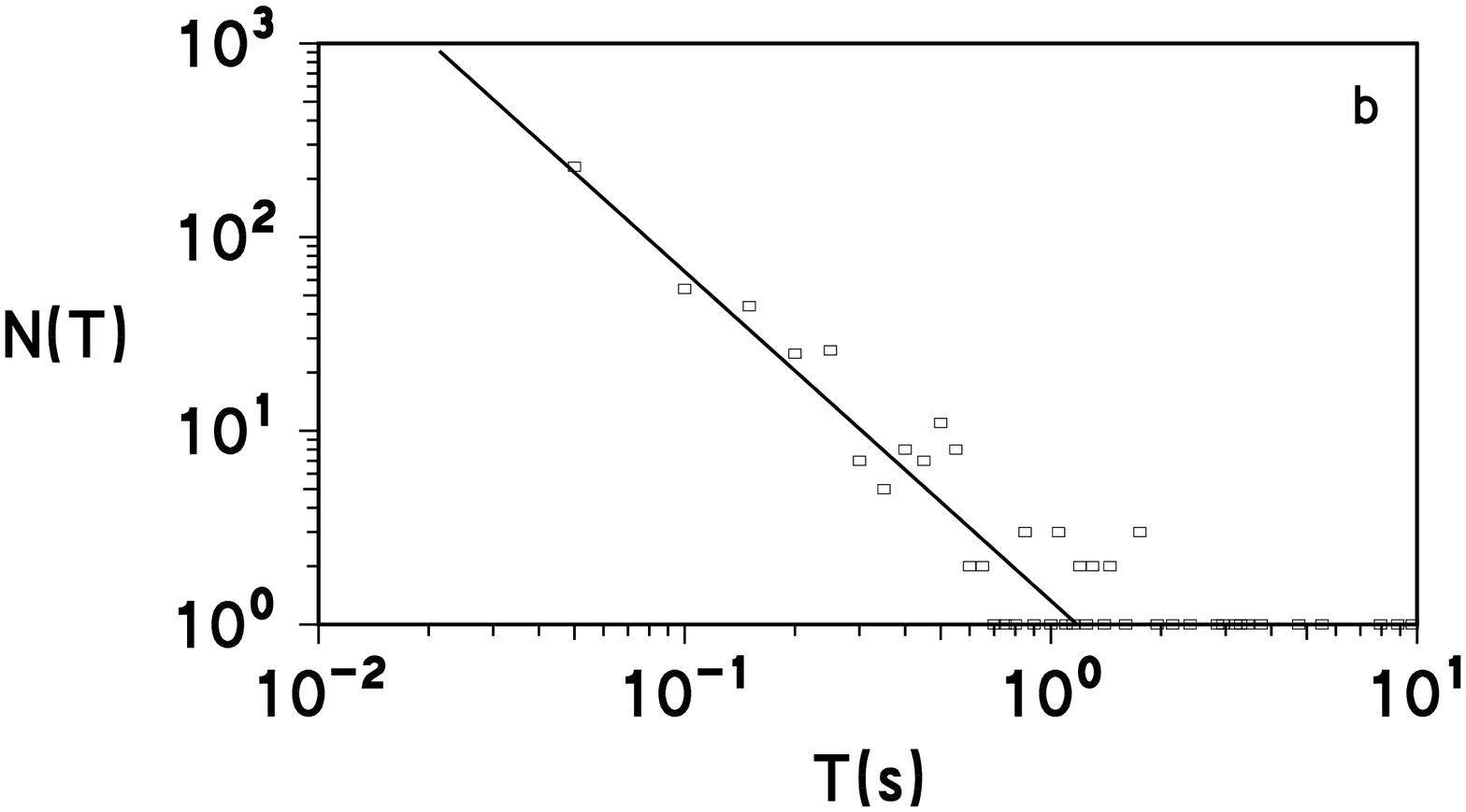}
\includegraphics[trim=50 100 20 100,width=0.40\textwidth,keepaspectratio]{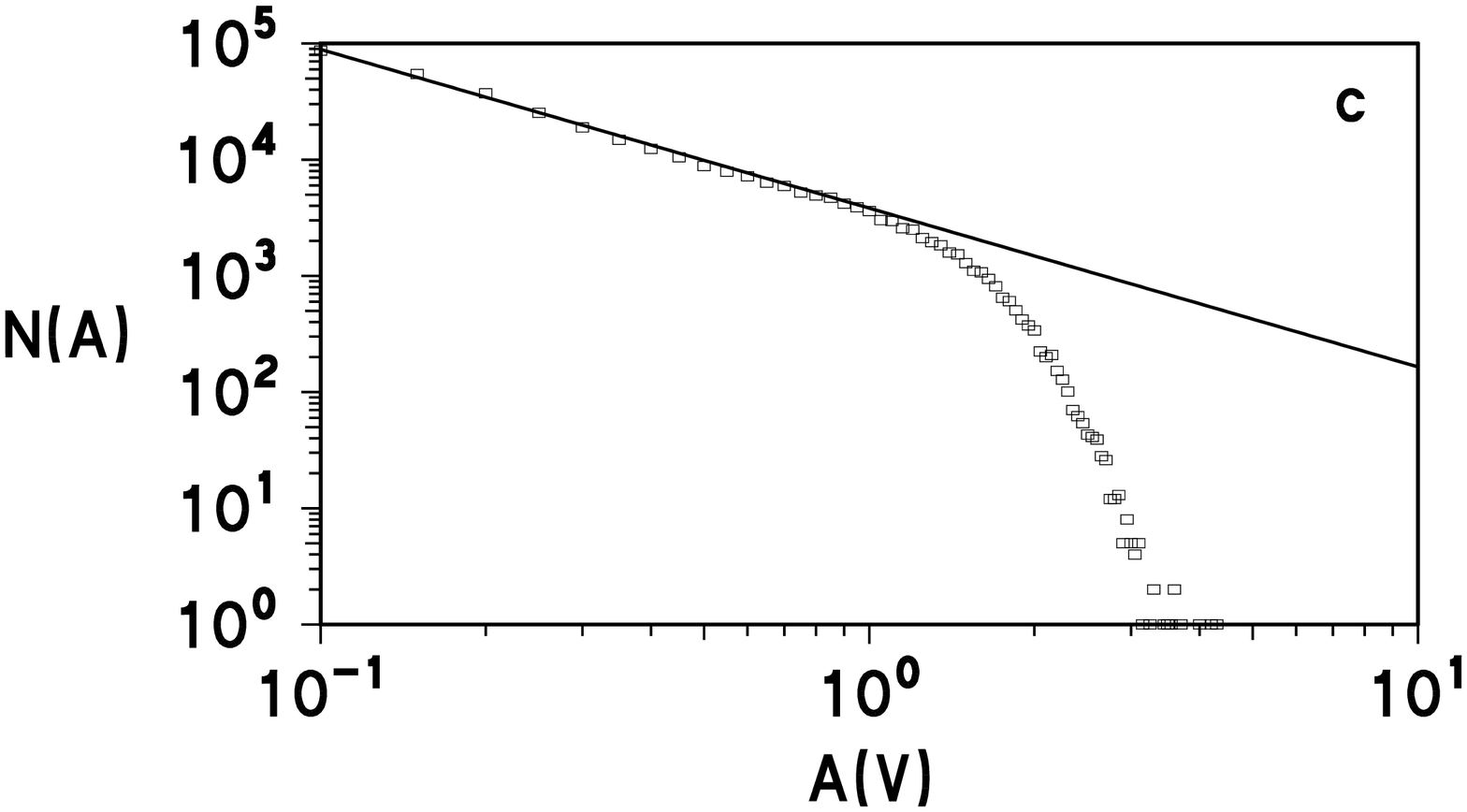}
\includegraphics[trim=50 100 20 100,width=0.40\textwidth,keepaspectratio]{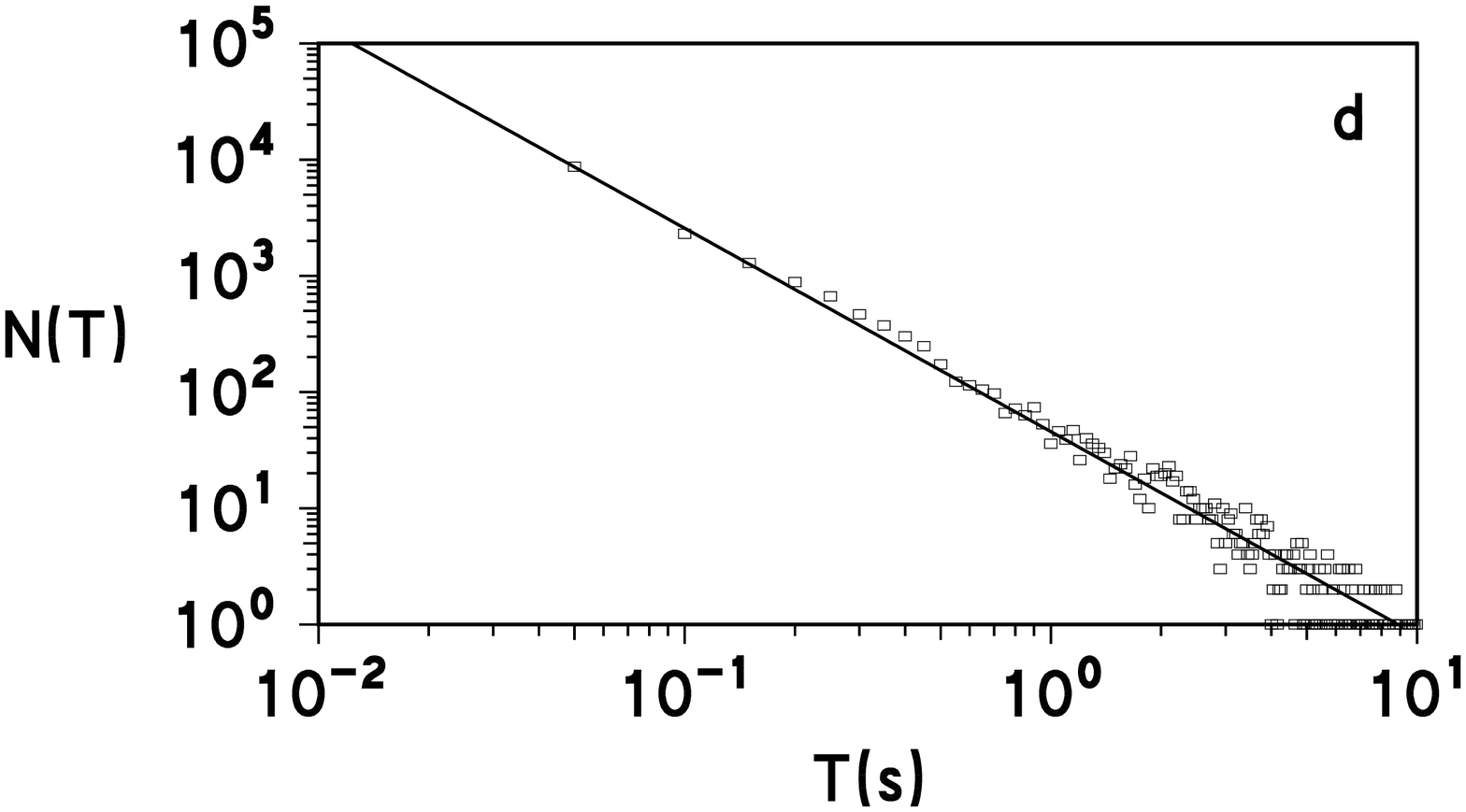}
\caption{Statistical distribution of sound intensity and power laws (1) and (2). 
a)``\textit{La calunnia \`{e} un venticello}'', amplitude distribution and law (2) with $\beta = 1.5$; b)``\textit{La calunnia \`{e} un venticello}'', time distribution and law (\ref{nt}) with $\gamma = 1.7$.
c)``\textit{Il Barbiere di Siviglia}'', amplitude distribution and law (\ref{na}) with $\beta = 1.3$; 
d) ``\textit{Il Barbiere di Siviglia}'', time distribution and law (\ref{nt}) with $\gamma = 1.7$.}
\label{f.1}
\end{center}
\end{figure}
However, to our knowledge up to now, nobody has done studies on the distribution of amplitude and time separation of sounds in the field of musical analysis.\par
These statistics were performed firstly in seismology. In 1894 Omori \cite{omori} showed that there is a temporal relationship between main shock and aftershocks, represented by the well known Omori's power law:
\begin{equation}
N(t)dt = Kt^{-\gamma} dt
\label{nt}
\end{equation}			 
where $K$ and $\gamma$ are constant and $N(t)dt$ is the number of events with time separation from the next one in the range $(t, t+dt)$.\par
In 1939 Ishimoto and Iida \cite{ii} found out a similar power law concerning the amplitude di\-stri\-bu\-tion of seis\-mic e\-vents. In 1954 the same law was expressed by Gutenberg and Richter in a different form \cite{gr}:
\begin{equation}
N(A)dA = HA^{-\beta} dA
\label{na}
\end{equation}					 
where $H$ and $b$ are constant and $N(A)dA$ is the number of earthquakes with amplitude between $A$ and $A+dA$.\par
Per Bak et al. \cite{bak} demonstrated that both these two power law distributions are characteristic of dynamic systems subjected to a ramification process: a perturbation generates instability in a point in the system and this instability starts a reaction chain causing events with no characteristic scale, leading up to catastrophic events that we call ``paroxysmal phases''. This kind of dynamics is present in a great variety of systems \cite{bak2}: economy (with its unforeseeable slump), ecosystems, earthquakes, paroxysmal phases of Acoustic Emission at Stromboli \cite{dmp}, phase transition in solids, Barkhausen effect, avalanches and, perhaps, the evolution of some diseases. It is present in information too: a news is born and, if it gains particular interest, propagates to many people just with a ramification process.\par
Moving this idea in the sphere of music, it is inevitable to think of Rossini's masterpiece ``\textit{La calunnia \`{e} un venticello\ldots}'' (``\textit{Calumny is a gentle wind\ldots}'') from ``\textit{Il Barbiere di Siviglia}''. In this aria words and music are in a special harmony: the avalanche-like evolution of calumny, that from little whispering becomes ``\textit{an earthquake, a thunderstorm, \ldots a cannon shot}'', has been set to a music whose acoustic intensity distribution, in amplitude and time (Fig. \ref{f.1} a and b), follows the same power law of earthquakes, of avalanches and avalanches of avalanches of ultrasonic bursts generated in the phase transitions \cite{dp} or in volcanic activity \cite{dmp}. Also the distribution relative to the whole opera follows the power laws (1) and (2) (Fig. \ref{f.1} c and d).\par
During the reproduction of the analyzed pieces, an electric signal from the loudspeaker was sent to a peak detector whose output level $A(t)$ quickly (within 0.5 ms) settled at the positive peak level of the signal, and after returned to zero with a 20 ms time constant. Then, fixed a suitable threshold ($\bar{A}$ = 100mV) (Fig. 2), the signals with maximum amplitude $A(t)>\bar{A}$, each of them we named event, were digitized by a PC with a frequency rate of 100 samples/s. \par
\begin{figure}[!tbh]
\begin{center}
\includegraphics[trim=20 420 20 50,width=0.37\textwidth,keepaspectratio]{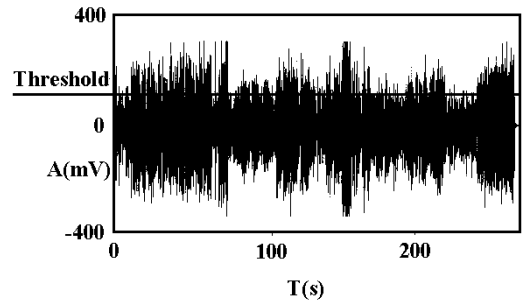}
\caption{Amplitude variation of one of the examinated pieces and threshold used to determinate the events.
}
\label{f.2}
\end{center}
\end{figure}
\begin{figure}[!tbh]
\begin{center}
\includegraphics[trim=0 50 0 50,width=0.45\textwidth,keepaspectratio]{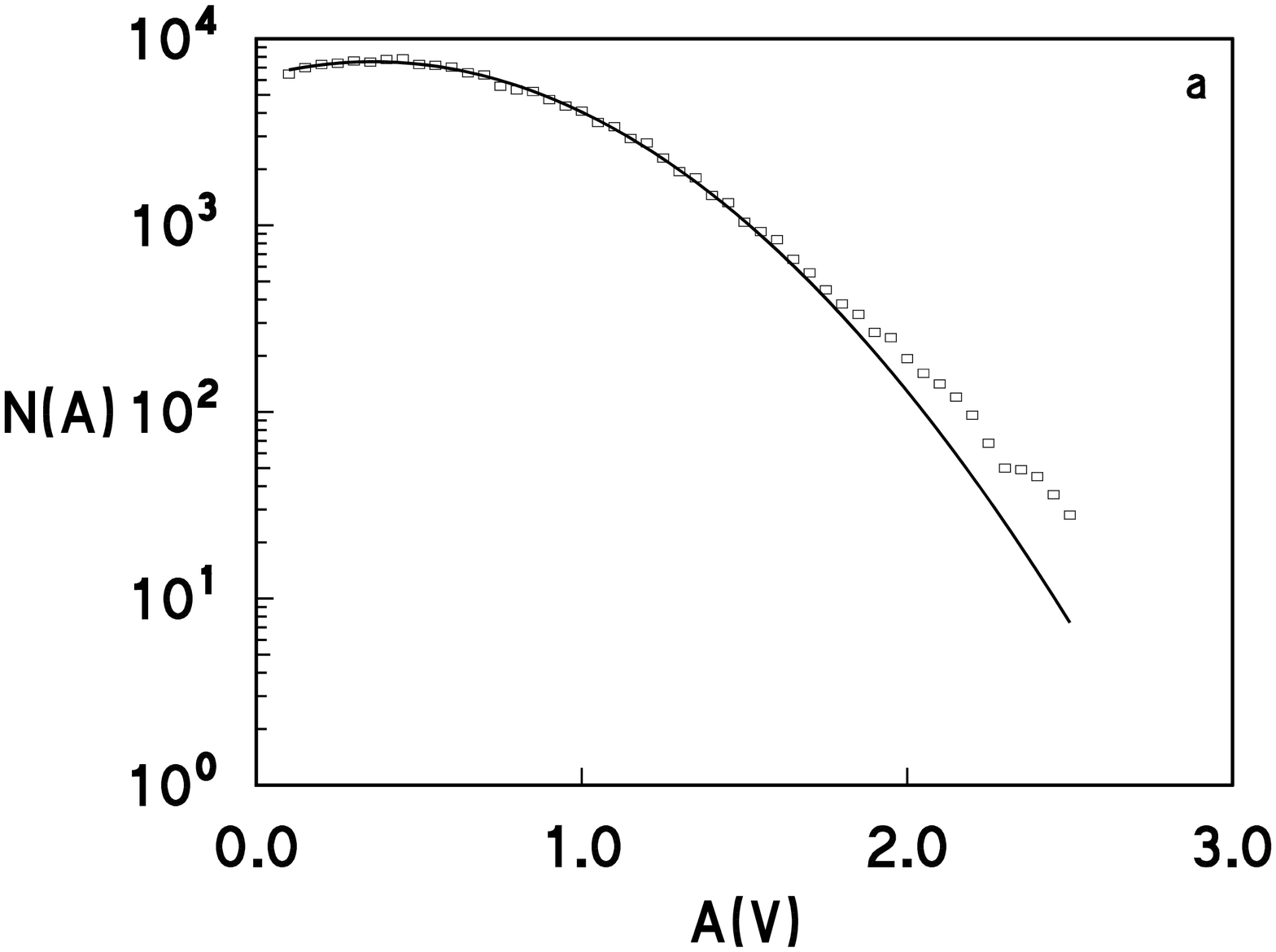}
\includegraphics[trim=0 50 0 50,width=0.45\textwidth,keepaspectratio]{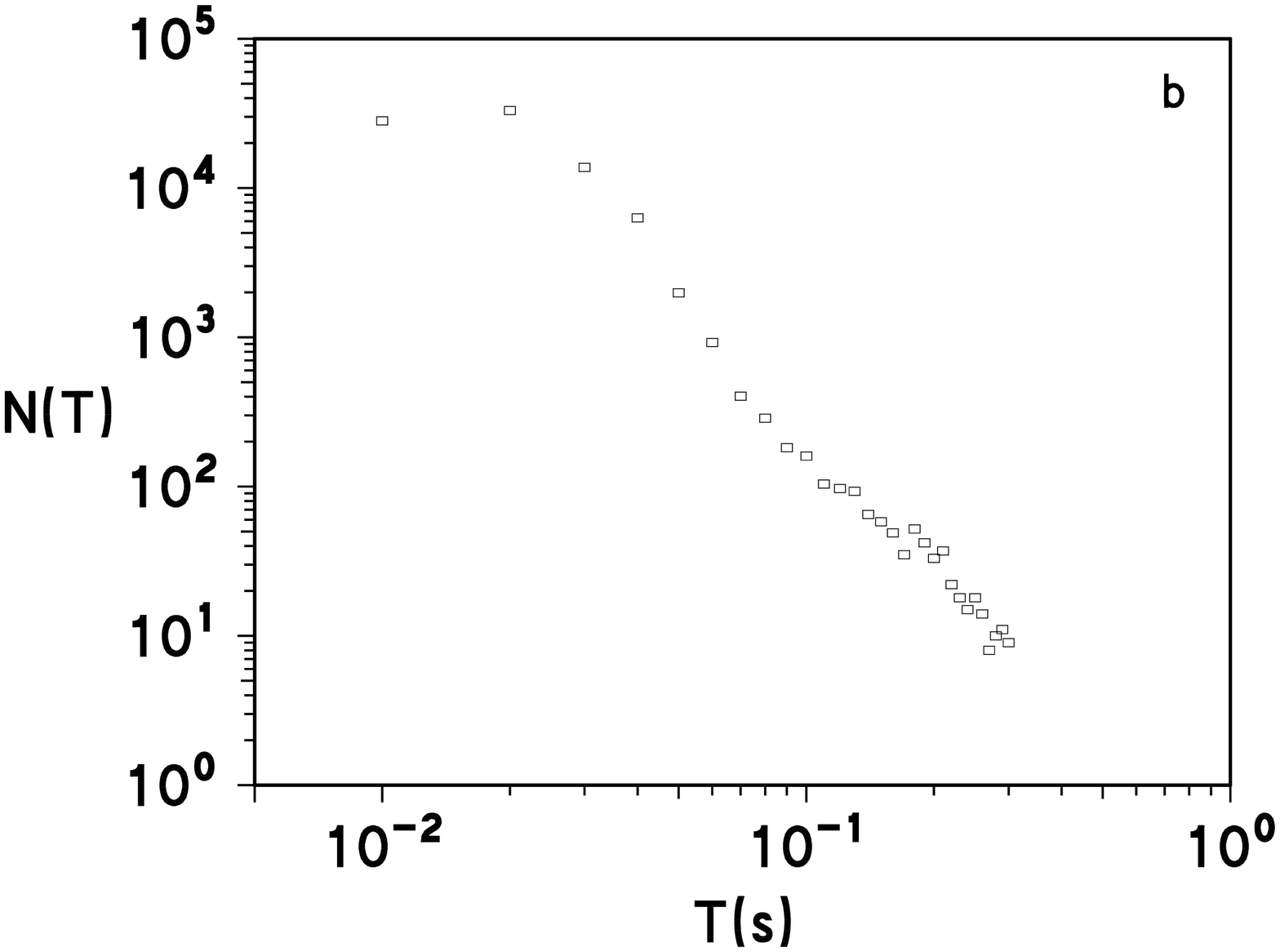}
\caption{Ella Fitzgerald and Louis Armstrong collection. Amplitude distribution of sound intensity (a), where the line represents a gaussian distribution, and (b) time distribution.}
\label{f.3}
\end{center}
\end{figure}
Finally we calculated amplitude and time distribution of the obtained time series of sound intensity.\par
\begin{figure}[!tbh]
\begin{center}
\includegraphics[trim=0 50 0 50,width=0.45\textwidth,keepaspectratio]{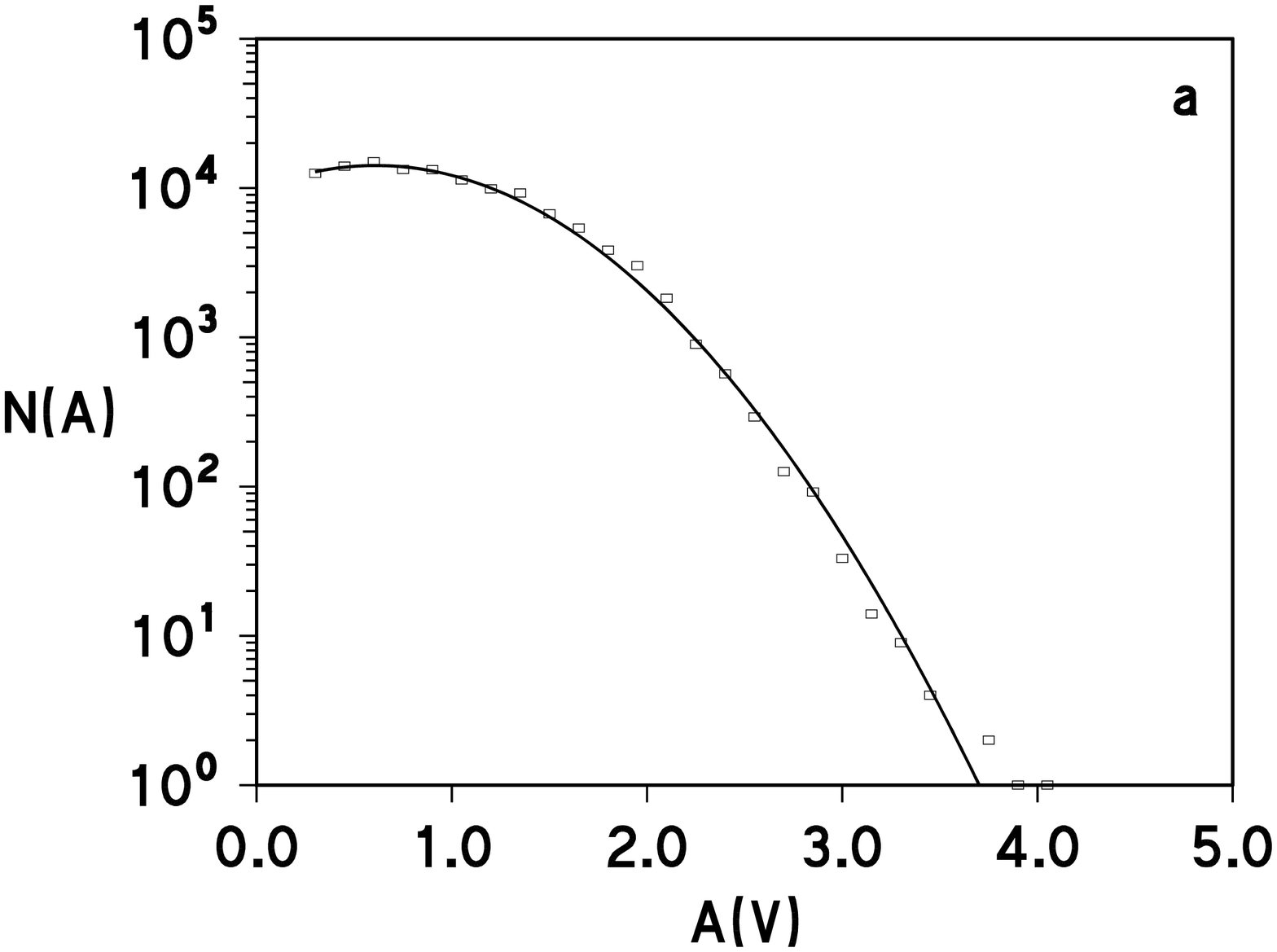}
\includegraphics[trim=0 50 0 50,width=0.45\textwidth,keepaspectratio]{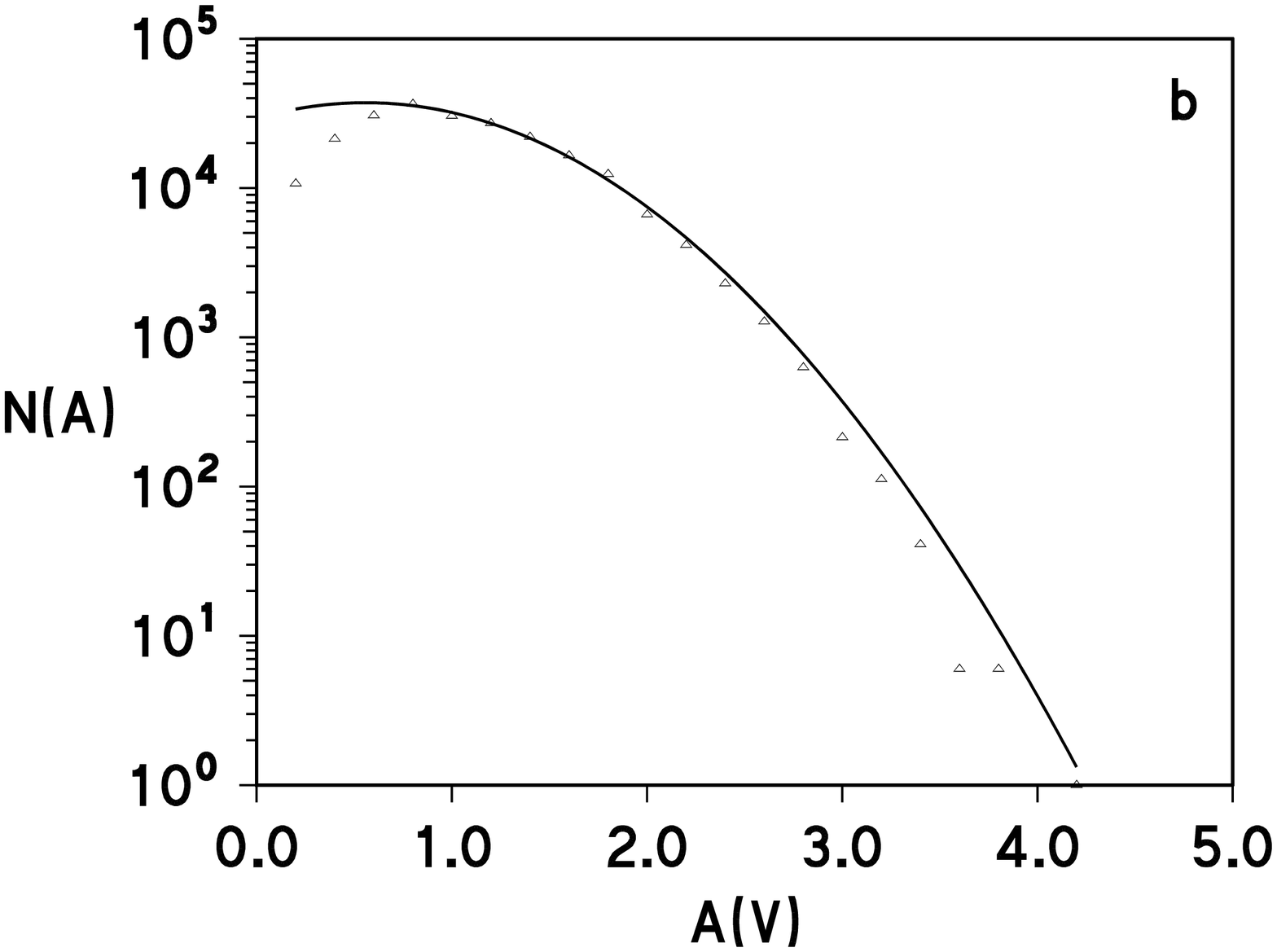}
\caption{Amplitude distribution of sound intensity of (a) ''A momentary lapse of reason`` by the Pink Floyd and (b)  ''Quadrophenia`` by The Who. The line represents a gaussian distribution.}
\label{f.4}
\end{center}
\end{figure}

In the same way, we analyzed other kind of music. Some examples are reported in Fig. \ref{f.3}, relative to a greatest hits collection of Ella Fitzgerald and Louis Armstrong, and in Fig. \ref{f.4} where only the amplitude distribution is plotted, being the time distribution  similar to that reported in Fig \ref{f.3} b, and not clearly characterized by a simple law. From these results, concerning just ten different authors, there seems to exist a remarkable distinction: jazz, pop and rock music give rise to gaussian distribution in amplitude, typical of noise, while pieces of classical authors are characterized by a more complex distribution law.
\section{Conclusion}
\label{sec:2}
We present these results just as a nice curiosity, but have no pretension at giving new perspective in the field of musical analysis. Perhaps it is possible that from further statistical studies, new classification methods of authors pieces and kinds of music can arise, even thought no distinction between different kind of music seems to emerge from spectral analysis \cite{voss1,voss2,boon,nett}.\par
For the present no general consideration can be drawn from our preliminary results. However, if new analysis shows the uselessness of our approach, it will remain a nice coincidence (but can we really call it a coincidence?) for the piece ``\textit{La calunnia\ldots}'', where words talking about ado propagation has been put to a music that is itself an example of dynamic evolution towards the catastrophe.
%
%
%
%
%
%

\end{document}